\documentclass[twocolumn,showpacs,preprintnumbers,amsmath,amssymb]{revtex4}

\usepackage{graphicx}
\usepackage{dcolumn}
\usepackage{bm}

\begin{document}

\preprint{APS/123-QED}

\title{Effect of Long-Range Interaction on the Critical Behavior
of Three-Dimensional Disordered Systems}

\author{S.V. Belim}
 \email{belim@univer.omsk.su}
\affiliation{%
Omsk State University, 55-a, pr. Mira, Omsk, Russia, 644077
\textbackslash\textbackslash
}%

\date{\today}

\begin{abstract}
A field-theoretical description of the behavior of a disordered
Ising system with long-range interaction is presented. The
description is performed in the two-loop approximation in three
dimensions using the Pade-Borel resummation technique. The
renormalization group equations are analyzed, and the fixed points
determining the critical behavior of the system are found. It is
shown that the effect of frozen structural defects on a system
with long-range interaction may cause a change in its critical
behavior or smearing of the phase transition.

\end{abstract}

\maketitle

The effect of long-range interaction, which at long distances is
described by the power law $r^{–D–\sigma}$ , was studied
analytically in terms of the å expansion [1–3] and numerically by
the Monte Carlo method [4–6] for two- and one-dimensional systems.
It was found that the effect of long-range interaction on the
critical behavior of Ising systems is considerable for $\sigma <2$.
 The study [7] carried out for a three-dimensional space in the
two-loop approximation confirmed the prediction of the $\varepsilon$-expansion
for a homogeneous system with long-range interaction.

According to the results of [8, 9], the introduction of frozen
impurities into a system leads to a change in its critical
behavior. In this connection, it is of interest to determine the
effect of frozen structural defects on the critical behavior of a
system with long-range interaction.

This paper describes the critical behavior of a three-dimensional
disordered system a long-range interaction for different values of
the parameter $\sigma$.

The Hamiltonian of a system with long-range interaction
has the form

\begin{eqnarray}
&&H_0=\frac 12\int d^Dq(\tau _0+q^\sigma)S_qS_{-q} +\frac 12\int
d^Dq\Delta \tau _qS_qS_{-q}\nonumber\\
&&+u_0\int d^DqS_{q1}S_{q2}S_{q3}S_{-q1-q2-q3}
\end{eqnarray}
where $S_q$ denotes the fluctuations of the order parameter,
$D$ is the space dimension, $\tau_0\sim|T-T_c|$, $T_c$ is the critical
temperature, $u_0$ is a positive constant, and $\Delta\tau_q$ is
the random field of impurities of the random temperature type.

The critical behavior strongly depends on the parameter $\sigma$,
which determines the rate of interaction decrease with distance.
According to [1], the effect of long-range interaction on the system
behavior is considerable for $0<\sigma<2$, while, for $\sigma \geq 2$,
the critical behavior is equivalent to that of a system with short-range
interaction. Therefore, only the case of $0<\sigma<2$ is considered below.

For a low impurity concentration, the distribution of
the random field $\Delta\tau_q$ can be considered as Gaussian and
described by the function
\begin{equation}
P[\Delta \tau ]=A\exp [ -\frac 1{\delta_0}\int \Delta \tau_q^2d^Dq],
\end{equation}
where $A$ is the normalization factor and $\delta_0$
is a positive constant proportional to the concentration of frozen
structural defects.

Applying the replica procedure for averaging over
random fields caused by the frozen structural defects,
we obtain the effective Hamiltonian of the system
\begin{eqnarray}
&&H_{R}=\frac 12\int d^Dq(\tau _0+q^\sigma)\sum\limits_{b=1}^mS^b
_qS^b_{-q}\\
&&-\frac{\delta _0}2\sum\limits_{b,c=1}^m\int d^Dq(
S^b_{q1}S^b_{q2})(S^c_{q3}S^c_{-q1-q2-q3}) \nonumber\\
&&+u_0\sum\limits_{b=1}^m\int d^DqS^b_{q1}
S^b_{q2}S^b_{q3}S^b_{-q1-q2-q3}\nonumber
\end{eqnarray}
The properties of the initial system can be obtained in the
limit of the number of replicas (transforms) $m\rightarrow 0$.

Applying the standard renormalization group procedure
based on the Feynman diagram technique [10]
with the propagator $G(\vec{k})=1/(\tau+|\vec{k}|^a)$,
we arrive at the expressions for the functions $\beta_u$,$\beta_\delta$,
$\gamma_\varphi$ and $\gamma_t$, which
determine the differential renormalization group equation:
\begin{eqnarray}
    \beta_u&=&-(2\sigma-D)u\Big[1-36uJ_0+24\delta J_0\\
    &&+1728\Big(2J_1-J_0^2-\frac29G\Big)u^2-\nonumber\\
    &-&2304(2J_1-J_0^2-\frac 16G)u\delta+672(2J_1-J_0^2-\frac23G)\delta ^2\Big],\nonumber\\
    \beta _\delta&=&-(2\sigma-D)\delta \Big[1-24uJ_0+16\delta J_0\nonumber\\
    &&+576(2J_1-J_0^2-\frac 23G_1)u^2-\nonumber\\
    &-&1152(2J_1-J_0^2-\frac13G)u\delta +352(2J_1-J_0^2-\frac 1{22}G)\delta ^2\Big],\nonumber\\
    \gamma_t&=&(2\sigma-D)\Big[-12uJ_0+4\delta J_0+288\Big(2J_1-J_0^2-\frac13G\Big)u^2\nonumber\\
    &-&288(2J_1-J_0^2-\frac 23G)u\delta +32(2J_1-J_0^2-\frac12G)\delta ^2\Big], \nonumber\\
    \gamma_\varphi&=&(2\sigma-D)64G(3u^2-3u\delta+\delta^2),\nonumber\\
    J_1&=&\int \frac{d^Dq     d^Dp}{(1+|\vec{q}|^\sigma)^2(1+|\vec{p}|^\sigma)}\nonumber\\
    &&\cdot\frac{1}{(1+|q^2+p^2+2\vec{p}\vec{q}|^{\sigma/2})},\nonumber\\
    J_0&=&\int \frac{d^Dq}{(1+|\vec{q}|^\sigma)^2},\nonumber\\
    G&=&-\frac{\partial}{\partial |\vec{k}|^\sigma}\int \frac{d^Dq d^Dp}
    {(1+|q^2+k^2+2\vec{k}\vec{q}|^\sigma)}\nonumber\\
    &&\cdot\frac{1}{(1+|\vec{p}|^a)(1+|q^2+p^2+2\vec{p}\vec{q}|^{\sigma/2})}\nonumber
\end{eqnarray}
Let us redetermine the effective interaction vertex:
\begin{equation}\label{vertex}
    v_1=u\cdot J_0,\ \ \ \ \  v_2=\delta\cdot J_0.
\end{equation}
As a result, we obtain the following expressions for the
functions $\beta$, $\gamma_\varphi$ and $\gamma_t$:
\begin{eqnarray}\label{beta}
    &&\beta_1=-(2\sigma-D)\Big[1-36v_1+24v_2\\
    &&+1728\Big(2\widetilde{J_1}-1-\frac29\widetilde{G}\Big)v_1^2\nonumber\\
    &&-2304(2\widetilde{J_1}-1-\frac 16\widetilde{G})v_1v_2+672(2\widetilde{J_1}-1-\frac23\widetilde{G})v_2^2\Big],\nonumber\\
    &&\beta _2=-(2\sigma-D)\delta \Big[1-24v_1+16v_2\nonumber\\
    &&+576(2\widetilde{J_1}-1-\frac 23\widetilde{G})v_1^2\nonumber\\
    &&-1152(2\widetilde{J_1}-1-\frac13\widetilde{G})v_1v_2+352(2\widetilde{J_1}-1-\frac 1{22}\widetilde{G})v_2^2\Big],\nonumber\\
    &&\gamma_t=(2\sigma-D)\Big[-12v_1+4v_2+288\Big(2\widetilde{J_1}-1-\frac13\widetilde{G}\Big)v_1^2\nonumber\\
    &&-288(2\widetilde{J_1}-1-\frac 23\widetilde{G})v_1v_2+32(2\widetilde{J_1}-1-\frac12\widetilde{G})v_2^2\Big], \nonumber\\
    &&\gamma_\varphi=(2\sigma-D)64\widetilde{G}(3v_1^2-3v_1v_2+v_2^2).\nonumber
\end{eqnarray}
Such a redetermination makes sense for $\sigma\leq D/2$. In this
case, $J_0$, $J_1$ and $G$ are divergent functions. Introducing
the cutoff parameter $\Lambda$ and considering the ratios $J_1/J_0^2$
and $G/J_0^2$ we obtain finite expressions in the limit $\Lambda\rightarrow\infty$.

The values of the integrals were determined numerically. For the case $a\leq D/2$,
a sequence of the values of $J_1/J_0^2$ and $G/J_0^2$ was constructed for
different $\Lambda$ and approximated to infinity.

The critical behavior is completely determined by
the stable fixed points of the renormalization group
transformation. These points can be found by equating
the $\beta$-functions to zero:
\begin{equation}\label{nep}
    \beta_1(v_1^*,v_2^*)=0, \ \ \ \ \beta_2(v_1^*,v_2^*)=0.
\end{equation}
The stability requirement for a fixed point reduces to
the condition that the eigenvalues $b_i$ of the matrix
\begin{equation}  \displaystyle
B_{i,j}=\frac{\partial\beta_i(v_1^*,v_2^*)}{\partial{v_j}}
\end{equation}
lie in the right half-plane of the complex plane.

The critical exponent $\nu$ characterizing the growth of the correlation
radius in the vicinity of the critical point ($(R_c\sim|T-T_c|^{-\nu})$)
is obtained from the relation
\begin{eqnarray}
  \nu=\frac1\sigma(1+\gamma_t)^{-1}.\nonumber
\end{eqnarray}

The Fisher exponent $\eta$ describing the behavior of the
correlation function in the vicinity of the critical point
in the wave-vector space ($(G\sim k^{2+\eta})$) is determined from
the scaling function $\gamma_\varphi$: $ \eta=2-\sigma+\gamma_\varphi(v_1^*,v_2^*)$.
The values of other critical exponents can be determined from the scaling
relations.

It is well known that the perturbation series expansions
are asymptotic, and the interaction vertices of the
order parameter fluctuations in the fluctuation region
are sufficiently large, so that Eqs. (7) applies. Therefore,
to extract the desired physical information from
the expressions derived above, the Padé–Borel method
generalized to the many-parameter case was used. The
corresponding direct and inverse Borel transformations
have the form
\begin{equation}
\begin{array}{rl} \displaystyle
  & f(v_1,v_2)=\sum\limits_{i_1,i_2}c_{i_1,i_2}v_1^{i_1}v_2^{i_2}=\int\limits_{0}^{\infty}e^{-t}F(v_1t,v_2t)dt,  \\
  & F(v_1,v_2)=\sum\limits_{i_1,i_2}\frac{\displaystyle c_{i_1,i_2}}{\displaystyle(i_1+i_2)!}v_1^{i_1}v_2^{i_2}.
\end{array}
\end{equation}
For the analytic continuation of the Borel transform of
a function, a series expansion in powers of the auxiliary
variable è is introduced:
\begin{equation}  \displaystyle
   {\tilde{F}}(v_1,v_2,\theta)=\sum\limits_{k=0}^{\infty}\theta^k\sum\limits_{i_1,i_2}\frac{\displaystyle c_{i_1,i_2}}{\displaystyle k!}v_1^{i_1}v_2^{i_2}\delta_{i_1+i_2,k}\  ,
\end{equation}
and the [L/M] Pade approximation is applied to this
series at the point $\theta=1$. This approach was proposed
and tested in [11] for describing the critical behavior of
systems characterized by several interaction vertices of
the order parameter fluctuations. The property
(revealed in [11-14]) that the system retains its symmetry
when using the Pade approximants in the variable $\theta$
is essential in the description of multivertex models.

The table shows the stable fixed points of the renormalization
group transformation and the eigenvalues of
the stability matrix at a fixed point for the parameter
values $1.5< \sigma \leq 1.9$ One can see that stable fixed points
exist in the physical region ($v_1^*,v_2^*>0$) only when the
long-range interaction parameter is a $\sigma\geq 1.8$. The calculations
showed that, for any $\sigma<1.8$, the stable points of
a three-dimensional impurity system are characterized
by a negative value of the vertex.

The calculation of the critical exponents provided the
following values:
\begin{eqnarray}
   &&\sigma=1.9,\ \ \ \nu=0.706786,\ \ \  \eta=0.1344048\\
   &&\sigma=1.8, \ \ \ \nu=0.732789,\ \ \ \eta=0.250978\nonumber
\end{eqnarray}
Thus, for a three-dimensional Ising system with
long-range interaction, the introduction of frozen structural
defects leads to a change in its critical behavior if
the long-range interaction parameter is $\sigma\geq1.8$ and to
smearing the phase transition if the parameter is $\sigma<1.8$.

\begin{table}
\begin{center}
\begin{tabular}{|c|c|c|c|c|c|c|} \hline
 $\sigma$ & $v_1^{*}$& $v_2^{*}$ & $b_1$    & $b_2$       \\
\hline
 1.5 &-0.395432 &  0.951135 & 84.530   & 59.517    \\
 1.6 &-0.227628 &  0.594810 & 45.302   & 32.575    \\
 1.7 &-0.045234 &  0.274890 & 13.235   & 3.915     \\
 1.8 & 0.064189 &  0.046878 & $0.626^*$& $0.626^*$ \\
 1.9 & 0.066557 &  0.040818 &$0.559^*$ & $0.559^*$ \\
 \hline
\end{tabular} \end{center} \end{table}
The work is supported by Russian Foundation for Basic Research N 04-02-16002.
\newpage
\def\baselinestretch{1.0}

\end{document}